\pdfoutput=1
\documentclass[a4paper]{article}

\usepackage{comment}

\usepackage{xurl}

\usepackage{hyperref}
\usepackage{graphicx}
\graphicspath{{./figures/}}

\usepackage{epsfig}
\usepackage{subcaption}
\usepackage{calc}
\usepackage{amssymb}
\usepackage{amstext}
\usepackage{amsmath}
\usepackage{amsthm}
\usepackage{pslatex}
\usepackage{apalike}
\usepackage{algorithm2e}

\title{The State of Disappearing Frameworks in 2023}
\author{Juho Vepsäläinen, Arto Hellas, and Petri Vuorimaa \and Aalto University}

\begin{document}
\maketitle



\abstract{Disappearing frameworks represent a new type of thinking for web development. In the current mainstream JavaScript frameworks, the focus has been on developer experience at the cost of user experience. Disappearing frameworks shift the focus by aiming to deliver as little, even zero, JavaScript to the client. In this paper, we look at the options available in the ecosystem in mid-2023 and characterize them in terms of functionality and features to provide a state-of-the-art view of the trend. We found that the frameworks rely heavily on compilers, often support progressive enhancement, and most of the time support static output. While solutions like Astro are UI library agnostic, others, such as Marko, are more opinionated.}



\vspace{10pt}

\textbf{The paper was accepted for WEBIST 2023 and the final version will be available in the conference proceedings. Note that this version has been edited to compile on arXiv and the final one is shorter due to a two-column layout.}

\section{\uppercase{Introduction}}
\label{sec:introduction}
Since its beginning in the 90s, the world wide web \cite{bernerslee1992} has become an enormous success. Although initially designed with the delivery of static websites in mind, over time, the web has allowed developers to build cross-platform applications with relative ease \cite{severance2012javascript}. The evolution of JavaScript has supported the transition from a website delivery platform to an application platform \cite{wirfs2020javascript}. With this transition, the demands for the platform have grown over time as the users expect more. 

The growing expectations have been intertwined with the evolution of web application development approaches. While traditional web relied on refreshing pages during operation, approaches such as single page applications (SPAs) have raised the bar in responsiveness and interactivity~\cite{severance2012javascript,woychowsky2006}. As shown by \cite{vepsalainen2023rise}, SPAs come with a cost of their own as they depend on client-side JavaScript and may be challenging to optimize for Search Engine Optimization (SEO) purposes. One particular challenge is the increasing amount of code shipped to the client~\cite{httparchivejs}. Newer approaches, such as disappearing frameworks, try to address these problem points.

To remedy the problems of SPAs, disappearing frameworks shift the focus on shipping a minimal amount of JavaScript to the client; hence the term disappearing \cite{vepsalainen2023rise}. Depending on the implementation, the way the framework disappears may differ, and it is an ongoing space of technical competition. The crux of disappearing frameworks is to take the best ideas from the early web -- delivering static content -- and combine them with the lessons learned from building SPAs, delivering as little as possible, even zero, JavaScript to the client.

This article surveys the currently available disappearing frameworks to understand possible ways to implement them and to create a map of the emerging space. The overarching research question of this article is as follows: \textit{Which disappearing frameworks exist in the ecosystem, and how can they be characterized in terms of functionality and features while considering their pros and cons?}

The framing responds to the question, ``What are the pros/cons of the
solutions from a developer and a user perspective relative to the incumbent approaches'' proposed in~\cite{vepsalainen2023rise}. This more specific question motivates considering the discovered frameworks against popular solutions to understand how they differ from mainstream ones.

To address the question, we first discuss disappearing frameworks in the context of earlier work in Section~\ref{sec:background}, before providing a technical review and comparison of them in Section~\ref{sec:comparison}. We discuss the findings in Section~\ref{sec:discussion} and conclude the findings in Section~\ref{sec:conclusion} while outlining future research directions.

\section{\uppercase{Background}}
\label{sec:background}
As described in \cite{vepsalainen2023rise}, the evolution of the web can be characterized through several phases: formation in the 90s, the rise of SPAs, and re-evaluation of current practices. Although a simplified view, this evolution provides a brief background for understanding the current developments. Early websites and applications could be characterized as multi-page applications (MPAs), which relied heavily on server logic triggered through navigation \cite{kaluza2018comparison}. In part motivated by a need to reduce loading times~\cite{nah2004study}, SPAs lifted the need to reload the whole page on each content change, improving user experience (UX) \cite{kaluza2018comparison} while also fundamentally changing and improving the developer experience \cite{vepsalainen2023rise}. In the current phase of re-evaluating development practices, solutions mixing the benefits of both approaches are being explored, and disappearing frameworks form one of the potential options.


\subsection{Libraries and frameworks}

Libraries and frameworks form one of the fundamental divisions in web development. Libraries such as \href{https://react.dev/}{React} aim to do a single thing well. At the same time, frameworks such as \href{https://angular.io/}{Angular} come with opinions, potentially increasing developers' productivity if they align with the framework's approach and work within its constraints. The border between a library and a framework is occasionally unclear, but this rough definition is sufficient for the present discussion.

To add complexity to the discussion, we also acknowledge the existence and emergence of meta-frameworks such as Astro. Meta-frameworks are headless as they do not rely on a specific user interface (UI) library but let the developer decide which one, or even many, to use. 

\subsection{Sprinkles architecture a.k.a. jQuery and friends}



\href{https://jquery.com/}{jQuery} (2006) is an early example of a successful JavaScript library that has wide usage to this day\cite{jqueryUsage2023}. jQuery was developed to address browser deficiencies in terms of ergonomics. jQuery could reduce twenty lines of standard DOM API code to a mere three through its chaining design while solving browser-incompatibility issues underneath \cite{bibeault2015jquery}. The style jQuery adopted could be characterized as ``Sprinkles architecture'', where JavaScript is sprinkled into the application to add interactivity while following the principle of progressive enhancement~\cite{champeon2003progressive}. The example below shows what jQuery declarations look like:



\begin{small}
\begin{verbatim}
 $(".selector").on("click", () => alert("hi"))
\end{verbatim}
\end{small}


The architecture pioneered by jQuery is still relevant, and most recently, several solutions, such as \href{https://alpinejs.dev/}{Alpine.js} (2019), \href{https://sidewind.js.org/}{Sidewind} (2019), and \href{https://htmx.org/}{htmx} (2020) have taken its ideas and moved it to HTML markup itself while still allowing JavaScript to be used when necessary. 


\subsection{Current web application development landscape}

React, Angular, and Vue.js dominate the current web application development landscape. Still, it is good to keep in mind that they hold only a small portion of the global market, highlighting that modern web applications are still a small subset of the whole web\footnote{E.g., the global market share of React is approximately 3.7\% \cite{vepsalainen2023rise}.}. Overall, these libraries and frameworks rely on components to allow the reuse and composition of code, leverage a templating solution (i.e., JSX) for component definition, and use a process called hydration to make the client-side code alive to the user by enabling event handlers and running component logic \cite{vepsalainen2023rise}.

Although the current technologies enable the creation of complex web-based experiences, it is not always clear what to use and when, as requirements tend to differ depending on the use case \cite{jason2019}. For instance, for a simple website, using a complete framework might be too much in terms of complexity. On top of this, there is a development-related cost to consider, as frameworks can take time to configure and learn. Furthermore, by definition, frameworks are collections of opinions; at times, the framework opinions might not match what is required, which leads to additional work -- frameworks can make complex tasks possible and possible tasks easy. Still, it is challenging to go against the inherent opinions of the frameworks.

Similarly, there are differences in the performance of frameworks that influence their suitability for specific tasks. As an example, \cite{ollila2022modern} explored the cost of rendering using contemporary frameworks and observed that the cost of React grows radically with the number of components, not to mention the amount of component code that must be loaded. In particular, leveraging a compiler can yield benefits when optimizing client-side performance \cite{ollila2022modern}.

\subsection{Islands architecture}

As pointed out by \cite{vepsalainen2023rise}, islands architecture can be considered as a stepping stone towards disappearing frameworks. The idea of islands architecture is to let the developer define which portions of a page are interactive while attaching a loading strategy to each interactive section; these areas are loaded only when they are needed (e.g., a user scrolls to a location on the page that was previously not visible), while not being loaded at all if not needed. 

Compared to loading and hydrating the entire page before it becomes accessible to the user, islands are an improvement for the users. Astro framework has popularized the approach, and it is good to recognize islands architecture as a recent development aiming at the same direction (2019) \cite{jason2020,patterns2022}. We cover one possible implementation when discussing Astro in detail in Section~\ref{subsection:astro}.

\begin{figure*}
    \centering
    \includegraphics[scale=0.25]{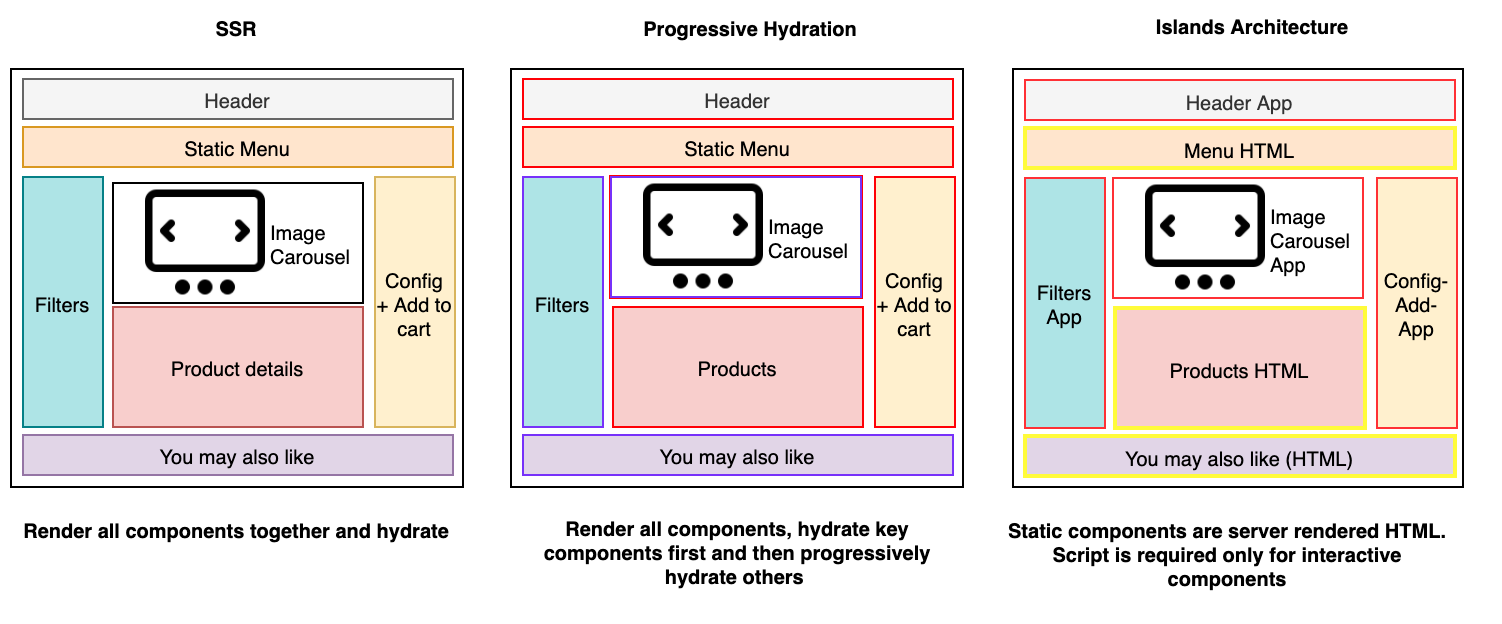}
    \caption{Three examples of rendering of a website. Server-side rendering (SSR) creates the website on the server, sending it to the client. Progressive hydration allows the prioritization of components rendered and shown to the client. Islands architecture, on the other hand, creates a static website on the server with placeholders for islands, which are then retrieved only if needed \cite{patterns2022}}
    \label{fig:islands-compared}
\end{figure*}

The key differences between server-side rendering (SSR), progressive hydration, and islands architecture are illustrated in Figure~\ref{fig:islands-compared}. Compared to regular hydration, progressive hydration goes further by hydrating key components first and the rest later \cite{patterns2022}.

\subsection{Disappearing frameworks}

The term disappearing frameworks appeared to the public in \cite{mediumDisappearingFrameworks} (2018), and as described by \cite{carniato2021}, disappearing frameworks represent a paradigm-level shift by rephrasing the problem addressed by contemporary frameworks. Rather than addressing the problem of developing complex web applications, disappearing frameworks aim to remove themselves from an application and try to start from close to zero cost in terms of JavaScript shipped to the client \cite{vepsalainen2023rise}. At the same time, disappearing frameworks leverage ideas, such as using components of the earlier generation, but how they are framed differs, especially from a loading point of view.

By shifting focus to the cost of what is shipped, disappearing frameworks reach towards the best practices discovered during the early web. Although the client still has to perform some work, the target is to defer the work performed and avoid it when possible. By reducing the amount of scripting in the client, accessibility is improved, especially in performance-limited contexts, such as mobile devices \cite{ollila2022modern}.


The idea of disappearing frameworks is consistent with Rich Harris' transitional web applications (TWAs) from 2021 that aim to capture the best ideas of both the traditional web and the SPAs \cite{rich2021} and practices such as progressive enhancement from 2008 \cite{alistapart2008} that encourage developers to think markup and styling first before applying JavaScript logic. TWAs go a step further, and the idea is that such a web application should be able to work without JavaScript enabled and, therefore, contain the necessary fallback mechanisms to work in this case.

Disappearing frameworks respond to the demands of both developers and users as they take the advancements gained during the SPA era and adapt them to the best practices discovered during the earlier practices established for developing MPAs. In other words, the movement aims to remedy the gap in the approaches while delivering better user experiences, especially in contexts with limited bandwidth and computing power. As a side benefit, the shift also aligns with green computing that highlights the need for increased mindfulness of resource usage~\cite{kurp2008green}.

\subsection{Developing web applications server-first}


While the previous examples have heavily focused on client-side development, there is also an emphasis on server-side functionality. For example, \href{https://www.phoenixframework.org/}{Phoenix} LiveView emphasizes what happens at the server, and sites written in LiveView can work without JavaScript, but when enabled, changes are relayed to the client through a WebSocket \cite{hexdocsPhoenixLiveViewx2014}. The approach allows \href{https://elixir-lang.org/}{Elixir} developers to build complex applications while staying within the Elixir language environment. The approach is interesting because it optimizes for First Meaningful Paint (FMP) while leaving JavaScript under the hood, allowing developers to stay within their preferred environment. Phoenix LiveView is not unique, and there are many implementations for other programming languages beyond Elixir, as listed in \cite{githubGitHubLiveviewsliveviews}.





\section{\uppercase{Technical Review}}
\label{sec:comparison}
By our definition, a framework is a disappearing framework if the focus is on aiming for zero or near zero cost in terms of JavaScript delivered to the client. The definition rules out older frameworks shipping a runtime and application code to evaluate at the client, giving a good baseline for evaluating newer options. In this article, we explicitly focus on frameworks that self-identify as ones that seek to minimize delivered JavaScript. To scope further, we limit our analysis to solutions that are under active development and have had a recent release. 

For the present analysis, we identified ten frameworks, outlined in Table~\ref{table:frameworks}, listing the projects and their characteristics while showing our selection of projects to study in detail. The identification of the frameworks was based on existing discussions (e.g.~\cite{mediumDisappearingFrameworks,devUnderstandingTransitional}) and an exploration of conferences focusing on modern web development practices (e.g., Future Frontend 2023 conference\footnote{\url{https://futurefrontend.com/2023/}}). 





\begin{table*}[h]
\caption{Potential web application frameworks and libraries}\label{table:frameworks} \centering
\begin{tabular}{|l|p{1.5cm}|p{2.6cm}|c|p{5.3cm}|c|}
  \hline
  Name & Type & Version & Compiled & Notes & Included \\
  \hline
  \href{https://svelte.dev/}{Svelte} & Library & 4.2.0, 2023-08-11 & \checkmark & Complemented by \href{https://kit.svelte.dev/}{SvelteKit} for routing and related functionality & \checkmark \\
  \hline
  \href{https://elderguide.com/tech/elderjs/}{Elder} & Framework & 1.7.5, 2022-05-31 & \checkmark & Built on top of Svelte & \\
  \hline
  \href{https://stenciljs.com/}{Stencil} & Library & 4.1.0, 2023-08-21 & & Focus on authoring Web Components & \\
  \hline
  \href{https://angular.io/}{Angular} & Framework & 16.2.3, 2023-08-30 & \checkmark & Conventional framework with support for Web Components & \\ 
  \hline
  \href{https://markojs.com/}{Marko} & Framework & 5.31.6, 2023-08-25 & \checkmark & Implements a DSL on top of HTML, supports streaming & \checkmark \\
  \hline
  \href{https://qwik.builder.io/}{Qwik} & Framework & 1.2.10, 2023-08-26 & \checkmark & Complemented by \href{https://qwik.builder.io/docs/qwikcity/}{Qwik City} for routing and related functionality & \checkmark \\
  \hline
  \href{https://astro.build/}{Astro} & Framework & 3.0.8, 2023-09-04 & \checkmark & Focus on islands architecture & \checkmark \\
  \hline
  \href{https://iles-docs.netlify.app/}{îles} & Static site generator & 0.9.5, 2023-04-07 & \checkmark & Implements partial hydration and comes with zero cost by default & \\
  \hline
  \href{https://slinkity.dev/}{Slinkity} & Framework & 1.0.0-canary.1, 2023-01-09 & & Based on \href{https://www.11ty.dev/}{11ty} static site generator and \href{https://vitejs.dev/}{Vite} bundler, early alpha & \\
  \hline
  \href{https://fresh.deno.dev/}{Fresh} & Framework & 1.4.2, 2023-08-17 & & Deno-based, edge friendly framework with SSR support & \checkmark \\
  \hline 
\end{tabular}
\end{table*}



\subsection{Astro} \label{subsection:astro}

Astro is a meta-framework built with islands architecture in mind. The project aims to allow developer experience (DX) familiar with contemporary frameworks within a static environment \cite{astrodocs}. Astro achieves this in several ways:


\begin{enumerate}
    \item Islands are supported out of the box - In other words, developers can decide which component boundaries should be interactive. \cite{astrodocs}
    \item Server-first API design - To minimize the cost of hydration, work is offloaded to the server as soon as possible. \cite{astrodocs}
    \item Zero JavaScript by default - Astro doesn't emit any JavaScript by default. \cite{astrodocs}
    \item Edge-ready - Astro sites can be deployed anywhere, edge included, due to its static nature. \cite{astrodocs}
    \item Customizable - Numerous extensions exist to expand the capabilities of Astro. \cite{astrodocs}
    \item UI-agnostic - Astro can host many contemporary JavaScript frameworks. \cite{astrodocs}
\end{enumerate}

\subsubsection{Astro islands}

Astro implements islands architecture through what it calls Astro islands \cite{astroAstroIslands}. The following example illustrates how to load a static React component through Astro: 

\begin{small}
\begin{verbatim}
---
import MyComponent from "../MyComponent.jsx";
---
<MyComponent />
\end{verbatim}
\end{small}

To make the component interactive, you have to mark it as an island while giving it a loading strategy as below \cite{astroAstroIslands}:


\begin{small}
\begin{verbatim}
---
import MyComponent from "../MyComponent.jsx";
---
<!-- The component is interactive on load -->
<MyComponent client:load />
\end{verbatim}
\end{small}

Beyond loading the island immediately, Astro provides other strategies, including \textbf{client:idle}, \textbf{client:visible}, and \textbf{client:media} to mention some \cite{astroTemplateDirectives}.






\subsubsection{Observations}

Astro is the first framework that embraced the islands architecture as a first-class citizen and used it as a part of their marketing effort. Astro is not bound to a specific UI library and provides flexibility in static and dynamic use cases by supporting both. Astro comes with zero cost by default for JavaScript shipped to the client, and the cost is added only by defining islands. It can be argued that what happens beyond that could be potentially costly. Still, at the same time, this could be seen as a pragmatic compromise as the approach allows leveraging what is available in the ecosystem of popular UI libraries such as React.

\subsection{Fresh}

Fresh calls itself the next-generation web framework, and it claims to have the following features \cite{freshFreshNextgen}: just-in-time rendering, island-based client hydration \cite{freshGentleIntroduction}, zero runtime overhead, no build step, no configuration, and TypeScript support. In other words, Fresh renders on demand over the edge (JIT) while supporting islands architecture, enabling developers to ship code for interactivity when needed. Due to its edge-oriented approach, it avoids the build step and configuration. By leveraging \href{https://deno.land/}{Deno} as the runtime instead of \href{https://nodejs.org/en}{Node.js} for its implementation, TypeScript support is gained out of the box.

Based on Fresh documentation, Fresh leverages \href{https://preactjs.com/}{Preact} and JSX for rendering \cite{freshIntroductionFresh}. Preact is a light (3 kB) implementation of \href{https://react.dev/}{React} API, making it an ideal choice for a framework like Fresh. The combination of Deno and Preact also restricts the project as it is not framework-agnostic like Astro. Still, at the same time, the choice is understandable, given the project constraints and focus.


\subsection{Svelte and SvelteKit}

According to its homepage, Svelte lets developers build ``cybernetically enhanced web apps'' \cite{svelteSveltex2022}. Svelte's central claims to fame are writing less code, lacking a virtual DOM, and being genuinely reactive. In technical terms, Svelte relies on a compiler-based approach, and compared to the incumbent frameworks, it claims to avoid the associated cost at the browser.

\subsubsection{Svelte compiler}

The description of Svelte aligns well with the idea behind disappearing frameworks. Furthermore, Svelte retains many features of the earlier solutions, including component orientation, and it comes with a templating language. At the same time, the hydration step is achieved through code generated by the Svelte compiler. The Svelte compiler accepts code such as the one below adapted from Svelte documentation \cite{svelteSveltex2022}:

\begin{small}
\begin{verbatim}
<script>
  let count = 0;
  function handleClick() { count += 1; }
</script>
<button on:click={handleClick}>
  Clicked {count} time{count > 1 ? "s" : ""}
</button>
\end{verbatim}
\end{small}

Although Svelte alone is enough for simple applications, it is often complemented by a solution such as Astro or SvelteKit. We already saw Astro at Subsection~\ref{subsection:astro} and will discuss SvelteKit next.

\subsubsection{SvelteKit}

SvelteKit is a framework that builds on top of Svelte and provides core features, such as routing and server-side rendering (SSR). SvelteKit leverages Vite bundler for the added functionality and implements developer-oriented features such as Hot Module Replacement (HMR) for adequate development flow \cite{sveltekit2022}.

While SvelteKit can run as a server in production mode, it also supports static site generation (SSG). A SvelteKit site can be hosted on top of popular edge platforms, including Cloudflare Pages, Netlify, and Vercel while allowing developers to adapt to any platform beyond the officially supported ones \cite{sveltekit2022}.

From the user's point of view, SvelteKit allows developing sites that work without JavaScript and that enhance progressively if JavaScript is available to provide higher quality user experience \cite{sveltekitFormActions}. SvelteKit provides flexibility to the developer by allowing them to decide how (CSR, SSR) and where to render a page (server, client) to support both dynamic and static use cases depending on what is required \cite{sveltekitPageOptions}.

\subsubsection{Observations}

Svelte and SvelteKit align well with the ideas behind disappearing frameworks. It is not a big surprise, given both projects were initiated by Rich Harris, the developer who introduced the idea of TWAs to the broader public. Given that TWAs are closely aligned with disappearing frameworks, it makes sense that Svelte and related solutions comply well with the target of shipping less JavaScript to the client.

There is some learning curve to Svelte's approach as it implies learning to use Svelte's DSL for defining components. The connection between state and template is mainly intuitive, but learning template control structures requires effort from the developer.

Svelte was one of the earliest JavaScript frameworks built as a compiler; other projects have followed that choice. Compiler-based approaches contrast earlier runtime and bundler-focused frameworks where client cost was not considered as crucial as disappearing frameworks.

\subsection{Qwik and Qwik City}

Qwik is a web framework built by Miško Hevery (Angular.js), Adam Bradley (Ionic), and Manu Almeida (Gin Framework) \cite{adservioQwikPostModern}. The framework addresses the conflicting requirements of interactivity and page speed \cite{devmQwiksMagic}. The conflict means you have to compromise in interactivity or speed due to the hydration cost of the current frameworks \cite{devmQwiksMagic}. According to Qwik documentation \cite{qwikdocs}, the framework is solving the quandary using the following means:

\begin{enumerate}
    \item Resumability over hydration -- Instead of hydrating, Qwik can resume code execution on demand.
    \item Automatic code splitting -- Qwik splits code aggressively due to the approach avoiding manual effort by developers. 
    \item Tiny runtime -- Qwik runtime is only one kilobyte.
    \item Compilation over runtime cost -- Qwik's optimizer pushes some state management to the server side. 
\end{enumerate}

It is these factors that make Qwik unique compared to its competition. Because of its approach, Qwik represents a paradigm-level shift in how to develop web applications. One of the ways it achieves its targets is by focusing on different metrics than its predecessors, namely Time to Interactive (TTI) over Time to Load (TTL).



While many frameworks focus on TTL, the core metric of Qwik is TTI \cite{infoworldIntroQwik}. The developers of Qwik are concerned about how soon a web page can become responsive to user interaction. This shift in perspective likely motivated the approach and the idea of resumability.

\subsubsection{Resumability}

Resumability allows pages to become interactive on demand based on user intent \cite{devmQwiksMagic} while picking up where the server left off \cite{adservioQwikPostModern}. In contrast to hydration-based approaches, there is less work to do for the browser as a part of it has already been performed. As a result, TTI can become low, and the page can quickly respond to user interaction.

The idea of resumability is not new, as illustrated by the example of jQuery (2006) \cite{devmQwiksMagic}. The difference with jQuery is that working with Qwik is similar to working with React regarding developer experience (DX). Qwik has adopted a similar component style and supports reactive state management out of the box \cite{devmQwiksMagic}. To leverage resumability, Qwik generates both server and client-side code while serializing using a so-called optimizer \cite{devmQwiksMagic}

\subsubsection{Optimizer and automatic code splitting}

The optimizer is a core part of Qwik generating code using automatic code splitting applied across component and event listeners, meaning Qwik defers loading necessary code as long as possible \cite{qwikresumability}. Given deferring isn't always the preferred behavior, Qwik supports standard preloading strategies \cite{qwikresumability}. The strategies are then used by a service worker set up by Qwik to monitor application state \cite{devmQwiksMagic}.

For code splitting, Qwik has been split architecturally into three parts \cite{devYourBundler}: view, state, and handlers. Usually, these three parts are coupled and held together in code. The coupling means there are three parts to download, parse, and execute together \cite{devYourBundler}. Even if only one part is needed based on user interaction, all three must be processed regardless \cite{devYourBundler}.

Because of the separation of concerns in Qwik, automatic code splitting has become possible at an unprecedented level. A specific dollar-based code convention is used to mark the code splitting boundaries, and then the optimizer can compile the code based on them \cite{qwikdocs}. Due to the shift in perspective, Qwik can achieve fine-grained code splitting out of the box that its predecessors couldn't due to their tighter coupling of concerns.

\subsubsection{Code splitting boundaries, reactive state, and components}

To better understand how code splitting boundaries work in Qwik, consider the counter-based example below from Qwik documentation \cite{qwikdocs}. It also illustrates Qwik's usage components and reactive state, specifically signals. 

\begin{small}
\begin{verbatim}
import * as qwik from "@builder.io/qwik";

export default qwik.component$(() => {
  const count = qwik.useSignal(0);
  return (
    <div>
      <span>Count: {count.value}</span>
      <button onClick$={() => count.value++}>
        Click
      </button>
    </div>
  );
});
\end{verbatim}
\end{small}

The example is relatively close to the code you would write in React. React created suitable programming interfaces as a pioneer, and Qwik decided to mimic them while adding twists on top.



\subsubsection{Qwik City}

Qwik City is a meta-framework comparable to Next.js for React \cite{devmQwiksMagic}. It provides the following features on top of Qwik \cite{devmQwiksMagic}: directory-based routing, nested layouts, file-based menus, breadcrumbs, support authoring content with .tsx or .mdx file formats, and data endpoints. The technical target of Qwik City is to provide the capabilities of an MPA with the benefits of a SPA \cite{devmQwiksMagic}. Therefore, navigation-wise, the solution avoids page refreshes commonly encountered in MPAs and allows UX familiar from SPAs.


\subsubsection{Observations}

Qwik represents a paradigm-level shift in how web applications are developed. It approaches the problem from a different angle. Compared to earlier solutions, it tries to quickly solve the issue of providing highly interactive pages to the client on a tooling level thanks to its compiler-based implementation. At the same time, it has adopted ergonomics familiar to developers from React, easing its adoption.

\subsection{Marko}

Marko's main claims are familiarity, performance, scalability, and trustworthiness. By familiarity, Marko means that it has been built on top of standard JavaScript, CSS, and HTML with tweaks as a DSL for HTML. Performance is achieved through streaming, partial hydration, optimizing the compiler, and a small runtime. Scalability is reached through component orientation, as the system can be expanded as required. Trustworthiness is gained by the fact that Marko powers high-traffic sites, such as ebay.com. \cite{markojsMarko}

\subsubsection{Marko DSL}

To illustrate Marko DSL, the documentation provides the following example showcasing how to loop through an array and map it as an HTML list \cite{markojsMarko}:

\begin{small}
\begin{verbatim}
<!doctype html>
<html>
<head><title>Hello Marko</title></head>
<body>
    <h1>My favorite colors</h1>
    <ul>
        <for|color| of=["red", "tan", "hue"]>
            <li style=`color:${color}`>
                ${color.toUpperCase()}
            </li>
        </for>
    </ul>
    <shared-footer/>
</body>
</html>
\end{verbatim}
\end{small}

\subsubsection{Observations}

Marko fits the definition of a disappearing framework well, as one of its design goals is a small runtime and optimized rendering. Furthermore, leveraging partial hydration and streaming further improves user experience as a page can be rendered while the user receives data. Marko's DSL has a learning curve that plugins for popular code editors have alleviated.



\section{\uppercase{Discussion}}
\label{sec:discussion}
Several frameworks have already adopted ideas suitable for disappearing frameworks. Many advertise to ship zero JavaScript to the client by default, which shows that the developers of the frameworks are aware of the problem of increasing delivered JavaScript~\cite{httparchivejs} and the associated costs.

\subsection{Main observations}

Our technical review highlights the following observations: many frameworks are built as compilers, many solutions support static output, and there are different takes on interoperability.






Compared to earlier frameworks, such as React, which leveraged a compiler only for its JSX templating, it seems the new generation relies heavily on compilers. The shift enables the new frameworks to keep the runtime shipped to the client as lean as possible. The shift is supported by \cite{ollila2022modern}, as he states that runtime-based approaches are costly and using a compiler helps with client-side performance.

Many solutions support static output, allowing them to work as SSGs \cite{petersen2016,camden2017}. For cases that do not support static output, a solution like \href{https://www.npmjs.com/package/ssr-to-html}{ssr-to-html} can be used to extract it, but at the same time, the results may not be ideal. 




Astro is an example of a UI library agnostic framework, while others enforce a specific approach \cite{astrodocs}. Qwik meets somewhere in the middle by leveraging JSX and providing compatibility with React \cite{builderQwikReact}. Marko is an extreme example with its custom DSL. 


\subsection{Methods for improving existing frameworks}

As replacing a framework can be costly, research seeks ways to optimize the use of existing frameworks. In \cite{vogel2023}, the authors introduce two new open-source frameworks that can delay JavaScript code without breaking it. Through these kinds of tools, it becomes possible to leverage modern development practices on aging codebases.




\subsection{Summary of results}


The rendering and hosting approaches, support for static output, and the stance on external UI libraries for the reviewed frameworks are outlined in Table~\ref{table:framework-observations}. 


\begin{table*}[h]
\caption{Comparison of disappearing frameworks}\label{table:framework-observations} \centering
\begin{tabular}{|l|p{8.6cm}|p{1.4cm}|p{2.4cm}|}
  \hline
  Name & Rendering and hosting approach & Static output & Stance on external UI libraries \\
  \hline
  \href{https://svelte.dev/}{Svelte} & Compiles only what is needed. Possible to host on popular edge platforms when using SvelteKit \cite{svelteAdaptersx2022}. & Through SvelteKit & Svelte only \\
  \hline
  \href{https://markojs.com/}{Marko} & Compiles what is needed and streams results to the client. Possible to host on Cloudflare Workers and platforms supporting Node.js \cite{markojsServerIntegrations}. & & Marko DSL only \\
  \hline
  \href{https://qwik.builder.io/}{Qwik} & Compiles features within split points loaded on demand and ships a minimal runtime for bootstrapping. Possible to host on edge and Node.js platforms \cite{builderDeploymentsQwik}. & \checkmark & Supports React \\
  \hline
  \href{https://astro.build/}{Astro} & Implements islands architecture and provides multiple strategies for loading the islands. It includes a compiler and supports rendering for dynamic use cases on the edge. & \checkmark & Works with React and others \\
  \hline
  \href{https://fresh.deno.dev/}{Fresh} & Renders on demand on top of the edge. Possible to host on Deno Deploy or through Docker \cite{freshDeploymentFresh}. & & Depends on Preact \\
  \hline 
\end{tabular}
\end{table*}


\section{\uppercase{Conclusion}}
\label{sec:conclusion}
Disappearing frameworks are a new trend in web development. In this paper, we addressed the research question \textit{Which disappearing frameworks exist in the ecosystem, and how can they be characterized in terms of functionality and features while considering their pros and cons?} by conducting a technical review of frameworks that comply with the definition of disappearing frameworks by aiming to delivery zero or near zero JavaScript to the client. We observe that many approaches rely on a compiler while also considering the state in which client-side JavaScript has been completely disabled. Several options also support static output, allowing easy hosting. In addition, solutions such as Astro provide limited interoperability with earlier UI libraries.



In this article, we addressed one of the questions proposed by \cite{vepsalainen2023rise} and gained a further understanding of the space of disappearing web frameworks. Additional research is needed to understand how the frameworks perform relative to the mainstream frameworks and each other. Performance studies should not be limited only to the performance experienced by the client, as developer experience can also significantly influence the adoption of new technologies, as hinted by \cite{ferreira2022adoption}. In addition, it would be worthwhile to evaluate the code authored using the frameworks in detail to understand differences in understandability, complexity, and the number of lines of code, for example.


\pagebreak

\bibliographystyle{apalike}
{\small
\bibliography{references}}

\begin{thebibliography}{}

\bibitem[Adservio, 2022]{adservioQwikPostModern}
Adservio (2022).
\newblock {Q}wik – {T}he {P}ost-{M}odern {F}ramework --- adservio.fr.
\newblock \url{https://www.adservio.fr/post/qwik-the-post-modern-framework}.
\newblock [Accessed 15-Nov-2022].

\bibitem[Astro, 2023]{astrodocs}
Astro (2023).
\newblock {G}etting {S}tarted --- docs.astro.build.
\newblock \url{https://docs.astro.build/en/getting-started/}.
\newblock [Accessed 13-Apr-2023].

\bibitem[Astro Directives, 2023]{astroTemplateDirectives}
Astro Directives (2023).
\newblock {T}emplate {D}irectives {R}eference --- docs.astro.build.
\newblock \url{https://docs.astro.build/en/reference/directives-reference/}.
\newblock [Accessed 17-Apr-2023].

\bibitem[Astro Islands, 2023]{astroAstroIslands}
Astro Islands (2023).
\newblock {A}stro {I}slands --- docs.astro.build.
\newblock \url{https://docs.astro.build/en/concepts/islands/}.
\newblock [Accessed 17-Apr-2023].

\bibitem[Berners-Lee et~al., 1992]{bernerslee1992}
Berners-Lee, T., Cailliau, R., Groff, J.-F., and Pollermann, B. (1992).
\newblock World-wide web: the information universe.
\newblock {\em Internet Research}.

\bibitem[Bibeault et~al., 2015]{bibeault2015jquery}
Bibeault, B., De~Rosa, A., and Katz, Y. (2015).
\newblock {\em jQuery in Action}.
\newblock Simon and Schuster.

\bibitem[Camden and Rinaldi, 2017]{camden2017}
Camden, R. and Rinaldi, B. (2017).
\newblock {\em Working with Static Sites: Bringing the Power of Simplicity to Modern Sites}.
\newblock " O'Reilly Media, Inc.".

\bibitem[Carniato, 2021a]{carniato2021}
Carniato, R. (2021a).
\newblock Understanding transitional javascript apps.
\newblock [Accessed 29-Sep-2022].

\bibitem[Carniato, 2021b]{devUnderstandingTransitional}
Carniato, R. (2021b).
\newblock {U}nderstanding {T}ransitional {J}ava{S}cript {A}pps --- dev.to.
\newblock \url{https://dev.to/this-is-learning/understanding-transitional-javascript-apps-27i2}.
\newblock [Accessed 13-Apr-2023].

\bibitem[Champeon, 2003]{champeon2003progressive}
Champeon, S. (2003).
\newblock Progressive enhancement and the future of web design.
\newblock \url{http://www.webmonkey.com/03/21/index3a.html}.
\newblock [Accessed over The Wayback Machine, 15-May-2023].

\bibitem[Ferreira et~al., 2022]{ferreira2022adoption}
Ferreira, F., Borges, H.~S., and Valente, M.~T. (2022).
\newblock On the (un-) adoption of javascript front-end frameworks.
\newblock {\em Software: Practice and Experience}, 52(4):947--966.

\bibitem[Fresh, 2023]{freshFreshNextgen}
Fresh (2023).
\newblock fresh - {T}he next-gen web framework. --- fresh.deno.dev.
\newblock \url{https://fresh.deno.dev/}.
\newblock [Accessed 19-Apr-2023].

\bibitem[Fresh Deployment, 2023]{freshDeploymentFresh}
Fresh Deployment (2023).
\newblock {D}eployment | fresh docs --- fresh.deno.dev.
\newblock \url{https://fresh.deno.dev/docs/concepts/deployment}.
\newblock [Accessed 19-Apr-2023].

\bibitem[Fresh Documentation, 2023]{freshIntroductionFresh}
Fresh Documentation (2023).
\newblock {I}ntroduction | fresh docs --- fresh.deno.dev.
\newblock \url{https://fresh.deno.dev/docs/introduction}.
\newblock [Accessed 19-Apr-2023].

\bibitem[GitHub LiveViews, 2023]{githubGitHubLiveviewsliveviews}
GitHub LiveViews (2023).
\newblock {G}it{H}ub - liveviews/liveviews: {P}hoenix {L}ive{V}iew workalikes for different languages and frameworks --- github.com.
\newblock \url{https://github.com/liveviews/liveviews}.
\newblock [Accessed 27-Apr-2023].

\bibitem[Gustafson et~al., 2008]{alistapart2008}
Gustafson, A., Overkamp, L., Brosset, P., Prater, S.~V., Wills, M., and PenzeyMoog, E. (2008).
\newblock Understanding progressive enhancement.
\newblock [Accessed 29-Sep-2022].

\bibitem[Hallie and Osmani, 2022]{patterns2022}
Hallie, L. and Osmani, A. (2022).
\newblock {I}slands {A}rchitecture --- patterns.dev.
\newblock \url{https://www.patterns.dev/posts/islands-architecture/}.
\newblock [Accessed 29-Sep-2022].

\bibitem[Harris, 2021]{rich2021}
Harris, R. (2021).
\newblock Have single-page apps ruined the web? | transitional apps with rich harris, nytimes.
\newblock [Accessed 29-Sep-2022].

\bibitem[Hevery, 2021]{devYourBundler}
Hevery, M. (2021).
\newblock {Y}our bundler is doing it wrong --- dev.to.
\newblock \url{https://dev.to/builderio/your-bundler-is-doing-it-wrong-ic0}.
\newblock [Accessed 14-Nov-2022].

\bibitem[{HTTP Archive}, 2023]{httparchivejs}
{HTTP Archive} (2023).
\newblock State of javascript.
\newblock \url{https://httparchive.org/reports/state-of-javascript}.
\newblock [Accessed 15-May-2023].

\bibitem[Jiang, 2023]{freshGentleIntroduction}
Jiang, A. (2023).
\newblock {A} {G}entle {I}ntroduction to {I}slands --- deno.com.
\newblock \url{https://deno.com/blog/intro-to-islands}.
\newblock [Accessed 27-Apr-2023].

\bibitem[Kalu{\v{z}}a et~al., 2018]{kaluza2018comparison}
Kalu{\v{z}}a, M., Troskot, K., and Vukeli{\'c}, B. (2018).
\newblock Comparison of front-end frameworks for web applications development.
\newblock {\em Zbornik Veleu{\v{c}}ili{\v{s}}ta u Rijeci}, 6(1):261--282.

\bibitem[Kurp, 2008]{kurp2008green}
Kurp, P. (2008).
\newblock Green computing.
\newblock {\em Communications of the ACM}, 51(10):11--13.

\bibitem[Marko, 2023]{markojsMarko}
Marko (2023).
\newblock {M}arko --- markojs.com.
\newblock \url{https://markojs.com/}.
\newblock [Accessed 13-Apr-2023].

\bibitem[Marko Server Integrations, 2023]{markojsServerIntegrations}
Marko Server Integrations (2023).
\newblock {S}erver {I}ntegrations | {M}arko --- markojs.com.
\newblock \url{https://markojs.com/docs/server-integrations-overview/}.
\newblock [Accessed 19-Apr-2023].

\bibitem[Miller, 2019]{jason2019}
Miller, J. (2019).
\newblock {A}pplication {H}olotypes: {A} {G}uide to {A}rchitecture {D}ecisions - {J}{A}{S}{O}{N} {F}ormat --- jasonformat.com.
\newblock \url{https://jasonformat.com/application-holotypes/}.
\newblock [Accessed 10-Jan-2023].

\bibitem[Miller, 2020]{jason2020}
Miller, J. (2020).
\newblock Islands architecture.
\newblock [Accessed 29-Sep-2022].

\bibitem[Nah, 2004]{nah2004study}
Nah, F. F.-H. (2004).
\newblock A study on tolerable waiting time: how long are web users willing to wait?
\newblock {\em Behaviour \& Information Technology}, 23(3):153--163.

\bibitem[Ollila et~al., 2022]{ollila2022modern}
Ollila, R., M{\"a}kitalo, N., and Mikkonen, T. (2022).
\newblock Modern web frameworks: A comparison of rendering performance.
\newblock {\em Journal of Web Engineering}.

\bibitem[O'Shaughnessy, 2018]{mediumDisappearingFrameworks}
O'Shaughnessy, P. (2018).
\newblock {D}isappearing {F}rameworks --- medium.com.
\newblock \url{https://medium.com/samsung-internet-dev/disappearing-frameworks-ed921f411c38}.
\newblock [Accessed 11-Apr-2023].

\bibitem[Petersen, 2016]{petersen2016}
Petersen, H. (2016).
\newblock From static and dynamic websites to static site generators.
\newblock {\em University of Tartu, Institute of Computer Science}.

\bibitem[Phoenix LiveView, 2023]{hexdocsPhoenixLiveViewx2014}
Phoenix LiveView (2023).
\newblock {P}hoenix.{L}ive{V}iew - {P}hoenix {L}ive{V}iew v0.18.18 --- hexdocs.pm.
\newblock \url{https://hexdocs.pm/phoenix_live_view/Phoenix.LiveView.html}.
\newblock [Accessed 27-Apr-2023].

\bibitem[Qwik, 2022a]{qwikdocs}
Qwik (2022a).
\newblock {O}verview - {Q}wik --- qwik.builder.io.
\newblock \url{https://qwik.builder.io/docs/overview/}.
\newblock [Accessed 14-Nov-2022].

\bibitem[Qwik, 2022b]{qwikresumability}
Qwik (2022b).
\newblock {R}esumable - {Q}wik --- qwik.builder.io.
\newblock \url{https://qwik.builder.io/docs/concepts/resumable/}.
\newblock [Accessed 14-Nov-2022].

\bibitem[Qwik Deployments, 2023]{builderDeploymentsQwik}
Qwik Deployments (2023).
\newblock {D}eployments - {Q}wik --- qwik.builder.io.
\newblock \url{https://qwik.builder.io/docs/deployments/}.
\newblock [Accessed 19-Apr-2023].

\bibitem[Qwik React, 2023]{builderQwikReact}
Qwik React (2023).
\newblock {Q}wik {R}eact - {Q}wik --- qwik.builder.io.
\newblock \url{https://qwik.builder.io/docs/integrations/react/}.
\newblock [Accessed 19-Apr-2023].

\bibitem[Qwik's magic, 2022]{devmQwiksMagic}
Qwik's magic (2022).
\newblock {Q}wik's magic is not in how fast it executes, but how good it is in avoiding doing any work --- devm.io.
\newblock \url{https://devm.io/javascript/qwik-javascript-hevery}.
\newblock [Accessed 15-Nov-2022].

\bibitem[Severance, 2012]{severance2012javascript}
Severance, C. (2012).
\newblock {J}ava{S}cript: Designing a language in 10 days.
\newblock {\em Computer}, 45(2):7--8.

\bibitem[Svelte, 2023]{svelteSveltex2022}
Svelte (2023).
\newblock {S}velte - {C}ybernetically enhanced web apps --- svelte.dev.
\newblock \url{https://svelte.dev/}.
\newblock [Accessed 13-Apr-2023].

\bibitem[Svelte Adapters, 2023]{svelteAdaptersx2022}
Svelte Adapters (2023).
\newblock {A}dapters - {D}ocs - {S}velte{K}it --- kit.svelte.dev.
\newblock \url{https://kit.svelte.dev/docs/adapters}.
\newblock [Accessed 19-Apr-2023].

\bibitem[SvelteKit, 2023]{sveltekit2022}
SvelteKit (2023).
\newblock {I}ntroduction - {D}ocs - {S}velte{K}it --- kit.svelte.dev.
\newblock \url{https://kit.svelte.dev/docs/introduction}.
\newblock [Accessed 13-Apr-2023].

\bibitem[SvelteKit Form Actions, 2023]{sveltekitFormActions}
SvelteKit Form Actions (2023).
\newblock {F}orm actions - {D}ocs - {S}velte{K}it --- kit.svelte.dev.
\newblock \url{https://kit.svelte.dev/docs/form-actions}.
\newblock [Accessed 13-Apr-2023].

\bibitem[SvelteKit Page Options, 2023]{sveltekitPageOptions}
SvelteKit Page Options (2023).
\newblock {P}age options - {D}ocs - {S}velte{K}it --- kit.svelte.dev.
\newblock \url{https://kit.svelte.dev/docs/page-options}.
\newblock [Accessed 13-Apr-2023].

\bibitem[Tyson, 2022]{infoworldIntroQwik}
Tyson, M. (2022).
\newblock {I}ntro to {Q}wik: {A} superfast {J}ava{S}cript framework --- infoworld.com.
\newblock \url{https://www.infoworld.com/article/3676577/intro-to-qwik-a-superfast-javascript-framework.html}.
\newblock [Accessed 14-Nov-2022].

\bibitem[Veps{\"a}l{\"a}inen et~al., 2023]{vepsalainen2023rise}
Veps{\"a}l{\"a}inen, J., Hellas, A., and Vuorimaa, P. (2023).
\newblock The rise of disappearing frameworks in web development.
\newblock In {\em International Conference on Web Engineering}, pages 319--326. Springer.

\bibitem[Vogel and Springer, 2023]{vogel2023}
Vogel, L. and Springer, T. (2023).
\newblock Waiter and autratac: Don't throw it away, just delay!
\newblock In Garrig{\'o}s, I., Murillo~Rodr{\'i}guez, J.~M., and Wimmer, M., editors, {\em Web Engineering}, pages 278--292, Cham. Springer Nature Switzerland.

\bibitem[w3techs, 2023]{jqueryUsage2023}
w3techs (2023).
\newblock {U}sage {S}tatistics and {M}arket {S}hare of j{Q}uery for {W}ebsites, {M}ay 2023 --- w3techs.com.
\newblock \url{https://w3techs.com/technologies/details/js-jquery}.
\newblock [Accessed 08-May-2023].

\bibitem[Wirfs-Brock and Eich, 2020]{wirfs2020javascript}
Wirfs-Brock, A. and Eich, B. (2020).
\newblock {J}ava{S}cript: the first 20 years.
\newblock {\em Proceedings of the ACM on Programming Languages}, 4(HOPL):1--189.

\bibitem[Woychowsky, 2006]{woychowsky2006}
Woychowsky, E. (2006).
\newblock {\em {A}{J}{A}{X}: Creating web pages with asynchronous {J}ava{S}cript and {X}{M}{L}}, volume~8.
\newblock Prentice Hall Upper Saddle River, NJ, USA.

\end{thebibliography}

\end{document}